\documentclass{llncs}
\usepackage{graphicx}
\usepackage{verbatim}
\usepackage{listings}
\usepackage{float}
\usepackage{color}
\usepackage{url}
\begin{document}
\bibliographystyle{splncs}

\title{Applying Constraint Solving to the Management of Distributed Applications}
\author{Andrew J. McCarthy, Alan Dearle and Graham Kirby}
\institute{School of Computer Science, University of St Andrews, Jack Cole Building, North Haugh, St Andrews, Fife, KY16 9SX \newline\email{\{ajm,al,graham\}@cs.st-andrews.ac.uk}}
\maketitle

\begin{abstract}
We present our approach for deploying and managing distributed component-based applications. A Desired State Description (DSD), written in a high-level declarative language, specifies requirements for a distributed application. Our infrastructure accepts a DSD as input, and from it automatically configures and deploys the distributed application. Subsequent violations of the original requirements are detected and, where possible, automatically rectified by reconfiguration and redeployment of the necessary application components. A constraint solving tool is used to plan deployments that meet the application requirements.
\end{abstract}

\section{Introduction}
Distributed applications are ubiquitous and are used to provide crucial capabilities in a wide range of fields. In commerce, business processes are increasingly manifested as distributed graphs of co-operating services. In the data centre, web applications are generally designed and deployed as distributed three-tier applications. Wireless sensor networks form distributed applications, used to measure and monitor various phenomena in a range of environments. Such distributed applications require considerable expertise and effort to design, build, deploy and manage, resulting in high costs.

All of the tasks in application development are challenging and complex, from deciding on the appropriate abstractions and the architectural design to the implementation and testing of the business logic. The development task is further complicated when the application is distributed and comprises many different co-operating components executing on different hosts. The developer must:

\begin{itemize}
\item{decide how the application should be partitioned into a set of components which will execute over a collection of hosts}
\item{provide mechanisms for binding components and services}
\item{consider the additional failure modes which are possible due to the application's distributed nature}
\end{itemize}

During development, an application is usually tested continuously. Unit testing of components may be straightforward. Integration testing of the application as a whole is often more challenging, since its distributed nature must be taken into account. The developer will usually need to fully deploy the application in a test setting before shipping it to a production environment.

After an application has been developed and tested it must be deployed. This involves choosing an execution site for each component, arranging delivery of code to each execution site, and orchestrating the instantiation of the distributed application as a graph of interdependent components.

Once deployed, an application must be managed. This requires that the application is monitored to ensure that it is operating as intended and that corrective action is taken when a failure occurs. There are many issues to consider when managing a distributed application; possible problems include loss of network connectivity, over-demand for network bandwidth or application services, failure of components, failure of computational resources and unbalanced resource utilization. For a system administrator, diagnosing such failures or conditions and creating a recovery plan is a complex, difficult and time-consuming task. Eventually, new functionality will be introduced into the application, perhaps in response to changing operating conditions or to fix a bug, requiring the implementation of the application to be upgraded. This may involve introducing new components, upgrading existing components, utilizing new hardware or provisioning for a changed level of demand for the application.

In this paper we present an approach which addresses the problems and issues described above. Our approach is based on \emph{Desired State Management}, a declarative approach to the problems of deploying, managing and maintaining a distributed application. An application administrator describes the properties of a \emph{desired state} for a distributed application; the runtime system then configures, deploys and manages the application in an attempt to ensure that it conforms to that desired state. An application may be maintained or evolved by rewriting appropriate parts of the desired state description, perhaps to utilize a new implementation of a component, and the runtime system makes the appropriate changes to the deployed application. We present a proof of concept of our approach by introducing our Java implementation, the Deladas Runtime.

\section{Related Work}
\label{sec:relatedwork}

A number of languages have been developed to describe software architectures, including \cite{RefWorks:5397,RefWorks:5588,RefWorks:5895}. Typical of these is Acme \cite{RefWorks:4409}, which is intended to fulfil three roles: to provide an architectural interchange format for design tools, to provide a foundation for the design of new tools and to support architectural modelling. The Acme language supports the description of components joined via connectors, providing a variety of communication styles. Components and connectors may be annotated with properties that specify various attributes. Acme also supports a logical formalism based on relations and constraints that permits computational or run-time behaviour to be associated with the description of architectures. Acme does not, however, support the deployment of systems from the architectural descriptions, nor does it express constraints on physical resources.

The SmartFrog framework \cite{smartfrog:introduction} is similar to this work in its motivation, to address the problems of describing, deploying and managing complex, distributed assemblies of software components. SmartFrog consists of a declarative language for describing component collections and component configuration parameters, and a runtime environment which activates and manages the components to deliver and maintain running systems. In SmartFrog each component transitions through life-cycle states in lock-step with all other components in the deployment. The SmartFrog life-cycle/service model is similar to \cite{didier:hierarchical} which also utilizes constraint-based deployment. However, they advocate propagative deployment and maintain global constraints through the use of a consensus algorithm.

Hein and Ritter \cite{RefWorks:6382} describe a model driven system for the evaluation of global constraints in a distributed system. In their work an application is developed as a collection of CORBA components. Their system introspects each of the components to discover their state and structure. This information is used to create a snapshot of the application which is checked for consistency against the model.

The system described in this paper is most like \cite{823353} which is an environment for a adapting distributed applications to heterogeneous environments. The framework relies on declarative specifications (like Deladas), a runtime environment called \emph{Smock} and an AI planner (described below). The specifications contain descriptions of components that specify the required and implemented interfaces and conditions that must hold for component instances. The \emph{Smock} runtime provides three distinct pieces of functionality: a proxy mechanism, wrapper components on each host and cache coherence mechanisms. The proxy mechanism causes requests for services to be sent to the AI planner which decides on the appropriate selection and placement of components. The wrapper abstracts over node specifics allowing (Java) components to be loaded onto nodes in a generic manner. The cache coherency mechanisms provide directory level cache coherence amongst nodes sharing replicated data.

Another paper by some of the same authors \cite{838496} focuses on the problem of initial placement, which they call the \emph{Component Placement Problem} (CPP). The authors describe an AI planning algorithm called \emph{Sekitei} for solving the CPP which is defined to have five elements: the network topology, the component deployment behaviour, the application framework, the link crossing behaviour and the CPP goal. The network topology is described as a set of links between nodes; the application is defined as sets of interfaces and components. The link crossing behaviour is defined by functions which describe properties of inter-machine links. Finally, the CPP goal describes the desired state of the system with respect to placement. The \emph{Sekitei} algorithm utilises AI planning techniques to reduce the size of the search space in order to achieve scalability. \emph{Sekitei} is implemented as a pluggable component of the \emph{Smock} framework described above.

\cite{DBLP:conf/aips/BlytheDGKAMV03} exploits a planner to generate workflows for Grid applications. The planner searches alternative deployment plans using heuristics to attempt to find high quality deployment solutions.

\section{Approach}
\label{sec:approach}
Our general approach is to apply declarative and generative techniques to the deployment and management of distributed component-based software. Our goal is for application administrators to specify their deployment and management requirements for a distributed application \emph{declaratively} and \emph{concisely} in a high-level domain specific language. From this description, appropriate infrastructural elements are generated which are used to configure, deploy, monitor and manage the administrator's distributed application.

\begin{figure}[!hbtp]
\begin{center}
\includegraphics[scale=.60]{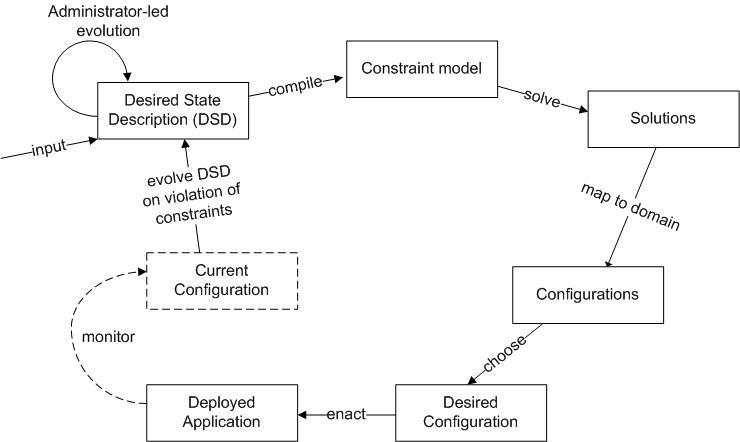}
\end{center}
\caption{Approach Overview.\label{fig:generalapproach}}
\end{figure}

In Figure~\ref{fig:generalapproach} we show an overview of our approach. The initial input to the process is a human readable \emph{desired state description (DSD)}. A DSD is a specification of the hardware and software resources available, plus a set of constraints on the deployed software, and optionally an optimization function expressed in terms of the desirable characteristics of the deployment. The constraints address aspects such as component connection topology, component instance placement on physical hosts, and choice of component implementations/versions.

The DSD is compiled into a lower-level domain-neutral constraint satisfaction problem (CSP). A CSP is expressed by declaring a set of variables whose values are drawn from a set of discrete domains, satisfying a set of given constraints. An example of a variable is an integer \emph{v1}, the value of which should be assigned from the continuous domain of \emph{0..10}. Constraints are typically binary relations, for example, the constraint that the value of one variable \emph{v1} is greater than the value of another variable \emph{v2}. Solving a CSP requires finding a set of consistent value assignments for the variables that satisfies all of the constraints. Each domain-neutral candidate solution is re-mapped into the specification domain to produce a set of candidate \emph{Configuration Description Documents} (CDDs). A CDD describes a particular mapping of components to hosts and interconnection topology that satisfies the DSD. The most suitable candidate configuration is chosen and \emph{enacted} to produce a \emph{deployment}.

When a configuration is enacted, \emph{bundles} are generated, encapsulating each component's implementation; these are sent to and instantiated on the appropriate hosts. Each bundle is an XML-encoded closure containing the code and data necessary to instantiate the component on a host. The bundle also includes custom probes and remote management interfaces to support monitoring and management of the deployed component. Implementations of these are generated as a side effect of the DSD compilation process described above. The management interfaces permit the manipulation of inter-component bindings during distributed application instantiation and maintenance.

As the application executes, it is monitored using the generated probes, which update a conceptually centralized (but physically distributed) model of the current configuration. This model is periodically compared to the desired state description. If it is detected that the model has ceased to satisfy the constraints specified in the DSD, the DSD is evolved to take account of the currently available hardware and software resources, and the solving cycle is repeated. This may occur as a result of failures of components or hosts within the deployment. Evolution may also be initiated by a human administrator, allowing components in the distributed application to be upgraded, or new components to be introduced.

\subsection{Specifying Desired State}
DSDs are specified in the declarative \emph{Deladas} language, similar in style to architecture description languages. The language's universe of discourse includes \emph{interfaces}, \emph{component types}, \emph{templates}, \emph{hosts} and \emph{constraint sets}.

An interface specifies the contract for interaction with a particular service. It does not describe individual methods as in languages such as CORBA IDL. Instead, it defines a URI specifying a concrete language-level interface, and the location of its implementation.

A component type describes a coarse-grain software entity. The definition includes the location of an implementation, and specifies the interfaces that it \emph{provides} to other components and \emph{requires} from other components. A component type may include a number of named \emph{properties}, either \emph{constant} with a defined value, or \emph{dynamic} with a value determined at execution time.

Templates are used to factor out details that are common to multiple component types. A template defines provided and required interfaces, and, optionally, properties. A component definition may extend a template. Templates allow abstract, reusable, component-based architectural patterns and styles \cite{garlan-exploiting} to be described. Such patterns and styles typically represent well understood, tried and tested architectures for component-based systems. Patterns and styles may also describe rules of a regulatory or compliance nature, for example, to restrict where certain classes of computation can take place.

A host definition describes a physical machine capable of executing components, with a number of constant or dynamic properties.

A constraint \cite{HCP-FM-06} set contains predicates defined over various characteristics of a deployed application. When all the predicates hold, the application is said to be \emph{compliant} with the constraint set, which constrains the manner in which the application is deployed---for example, with regard to the mapping of components to hosts, or the inter-component connection topology. Constraint sets are used to yield an initial satisfactory configuration for deployment, and also to detect desired state violations by a deployed application. The constraint algebra includes universal and existential quantifiers over components, templates and hosts, predicates over component connection topology, and expressions over properties.

We now introduce the syntax and expressibility of the Deladas language through a simple example application, a \emph{Maths} service, which allows users to invoke simple mathematical operations. The service is implemented by one or more equivalent components, each of which in turn requires references to other components that provide multiplication and division services. Once the application has been deployed, a user can access it by locating one of the main service components and invoking its interface. Figure~\ref{fig:mathsServiceDecl} shows an example DSD that describes the desired structure of the application.

\begin{figure}[!hbtp]
\begin{verbatim}
interface IMathsService (
  type = "java"
  specification = "com.math.IMathsService"
  implementation = "http://www.cs.st-andrews.ac.uk/deladas/mathsService.jar"
)

interface IMultiplicationService (
  type = "java" 
  specification = "com.math.IMultiplicationService"
  implementation = "http://www.cs.st-andrews.ac.uk/deladas/multiplicationService.jar"
)

interface IAdditionService (
  type = "java"
  specification = "com.math.IAdditionService"
  implementation = "http://www.cs.st-andrews.ac.uk/deladas/additionService.jar"
)

template MathsServiceTemplate (
  provides interface IMathsService
  requires IMultiplicationService multiplication, IAdditionService addition
  properties (
    constant string vendor
    dynamic int queriesPerSecond
  )
)

component type MathsService extends MathsServiceTemplate (
  implementation "http://www.cs.st-andrews.ac.uk/deladas/mathsService.jar"
  instantiate mathsServiceImpl with com.math.MathsService("hello")
  satisfy IMathsService using mathsServiceImpl  
  bind multiplication with mathsServiceImpl.setMultiplyService()
  bind addition with mathsServiceImpl.setAdditionService()
  initialise mathsServiceImpl.init()
  destroy mathsServiceImpl.shutdown()
  properties (
    vendor = "CalculusSoftware"
    queriesPerSecond providedBy mathsServiceImpl.qps()
    accuracy = 2
  )
)

component type MultiplicationService (
  provides interface IMultiplicationService
  implementation  "http://www.cs.st-andrews.ac.uk/deladas/multiplicationService.jar"
  instantiate multServiceImpl with com.math.MultiplicationService()
  satisfy IMultiplicationService using multServiceImpl
)

component type AdditionService (
  provides interface IAdditionService
  implementation "http://www.cs.st-andrews.ac.uk/deladas/additionService.jar"
  instantiate addServiceImpl with com.math.AdditionService()
  satisfy IAdditionService using addServiceImpl
)
\end{verbatim}
\begin{center}
\caption{Interface, Template and Component Declarations.\label{fig:mathsServiceDecl}}
\end{center}
\end{figure}

\subsubsection{Interfaces}
Three interfaces are declared: \emph{IMathsService}, \emph{IMultiplicationService} and \emph{IAdditionService}. The implementation type of each interface is Java; each specifies a concrete interface in the form of a Java interface name, and the location of a definition of that interface.

\subsubsection{Templates} 
The template \emph{MathsServiceTemplate} is introduced to illustrate the language construct, though it is not strictly necessary for this example. The \emph{provides} and \emph{requires} clauses state that the interface \emph{IMathsService} is provided to clients, and that the interfaces \emph{IMultiplicationService} and \emph{IAdditionService} are required by the template. The template also declares the properties \emph{vendor} and \emph{queriesPerSecond}.

\subsubsection{Component types} 
\label{sec:component-type-decls}

The component type \emph{MathsService} extends \emph{MathsServiceTemplate}, with the following consequences:

\begin{itemize}
\item{The component must \emph{satisfy} each of the provided interfaces declared by the template, in this case only \emph{IMathsService}. This means that the component type must specify how the provided interfaces are implemented.}
\item{The component will be provided with each of the required interfaces declared by the template, in this case \emph{IMultiplicationService} and \emph{IAdditionService}.}
\item{The component type must specify how values are associated with each of the properties declared in the template, in this case \emph{vendor} and \emph{queriesPerSecond}.}
\end{itemize}

The first line in the body of the component type declaration specifies the location of a Java implementation. This implementation is consulted during \emph{compilation} of the DSD to facilitate type checking and generation of appropriate glue code. Next, an \emph{instantiate} construct specifies that each instance of the component type should instantiate the implementation class \emph{com.math.MathsService} using the constructor taking the single given string parameter. The implementation object is referenced as \emph{mathsServiceImpl} in the remainder of the declaration. The next line specifies that the provided interface \emph{IMathsService} should be implemented by that implementation object. The following two lines describe how the component binds to its required interfaces. The first \emph{bind} clause states that the interface \emph{multiplication}, which is inherited from \emph{MathsServiceTemplate}, should be bound to the implementation object by calling its method \mbox{\emph{setMultiplyService()}}.

The \emph{initialise} clause specifies that when a component is initialised the method \emph{init()} should be called on the implementation object. The \emph{destroy} construct specifies that the method \emph{shutdown()} should be called on the implementation object before the component is destroyed. Finally, the component's properties are defined. Constant values are specified for \emph{vendor} and \emph{accuracy}, while the value of \emph{queriesPerSecond} is calculated dynamically by calling the method \emph{qps()} on the implementation object.

Figure~\ref{fig:mathsServiceConstraints} continues the DSD specification for the \emph{Maths} service. This part of the DSD defines the available hosts and a constraint set.
\begin{figure}[!ht]
\begin{verbatim}
host template Blade (speed = 1000)
host template CloudBlade (speed = 3000)

host h1 extends CloudBlade (address = "server5.deladas.com")
host h2 extends CloudBlade (address = "server6.deladas.com")
...
host h10 extends Blade (address = "server14.deladas.com")

constraintSet mathsServiceCons (
  forall MathsService mathsComponent in deployment (
    getHost(mathsComponent).speed >= 2000
  )
  and
  forall AdditionService additionComponent in deployment (
    card(connections(additionComponent.IAdditionService)) <= 2
  )
  and
  forall host h in deployment (card(getComponents(h)) <= 1)
  and
  card(instancesOf(MathsService in deployment)) >= 3
)
\end{verbatim}
\begin{center}
\caption{Constraints for the Maths Service.\label{fig:mathsServiceConstraints}}
\end{center}
\end{figure}

This example illustrates the use of \emph{host templates} to capture properties common to a set of hosts. The host templates \emph{Blade} and \emph{CloudBlade} define different values for the host property \emph{speed}. Five instances of each host template are  declared. The constraints specify that:

\begin{itemize}
\item{every component of type \emph{MathsService} must be located on a host with a speed of at least 2000.}
\item{no component of type \emph{AdditionService} may have more than two other components connected to it.}
\item{no host may have more than one component located on it.}
\item{there must be at least three components of type \emph{MathsService}.}
\end{itemize}

\section{Implementation}
\label{sec:implementation}
\label{sec:deladasruntime}
The Deladas Runtime is our prototype Java implementation of this approach. It builds on the CINGAL deployment framework~\cite{cingal:architecture,cingal:framework} and our initial work using constraints as a tool for autonomic management~\cite{deladas:introduction}. 

\subsection{Implementation of Compilation and Solving Mechanisms}
\label{sec:constraintsImpl}
A DSD contains the specification of the hardware and software resources available, plus a set of high-level constraints on the deployed application. The DSD is \emph{compiled} into a lower-level domain-neutral constraint satisfaction problem (CSP), which is then solved to yield a number of domain-neutral solutions. These solutions are then mapped back to the DSD domain to yield a number of candidate \emph{configurations}.

A constraint solver comprises a notation for modeling a CSP and one or more algorithms, such as backtracking and constraint propagation, that are used to solve the CSP. We have built a framework for compiling and solving DSDs built on ILOG's \emph{JSolver} \cite{RefWorks:6386}. Other available constraint solvers include \emph{MINION} \cite{RefWorks:6385} and \emph{Cream} \cite{RefWorks:6387}. \emph{JSolver} is a Java library that provides an object model for modeling CSPs and a collection of solving algorithms. A CSP is modeled by constructing a graph of Java objects in which each node is a constrained integer or boolean variable, and the arcs are constraints over those variables. Constraints include basic binary constraints and complex constraints over collections of variables, such as \emph{all different}. The solving process performs a search over the graph and assigns a values to each variable that is consistent with the specified constraints.

There is a wide gap between the level of abstraction used to model CSPs in tools such as \emph{JSolver} and the abstractions used by an application administrator to express desired state descriptions in Deladas. Similarly, the result of solving a CSP, namely a set of assignments to integer and boolean variables, is far removed from the architectural configurations required for application deployment. Thus, a DSD must be compiled into a CSP, and a CSP solution must be mapped back into a deployment configuration.

Our solving framework has three main parts:
\begin{itemize}
\item{A single general CSP, which models the general class of problem described by DSDs:}
\begin{itemize}
\item{the number of instances of each component type to be instantiated}
\item{the mapping of component instances to physical hosts}
\item{the interconnection topology between component instances}
\end{itemize}
\item{A \emph{CSP generator}, which generates a new CSP for each DSD, based on the general CSP.}
\item{An API providing a bridge between the constrained variables in the general CSP and the constraints expressed in a DSD.}
\end{itemize}
\subsubsection{General CSP}
The general CSP models the common parts of the configuration problem that must be solved for every DSD. It consists of a set of constrained variables and a set of default constraints over these variables. The CSP is the union of two sub-problems; the first models the number of instances of each component type and the hosts on which they are placed; the second models the component connection topology. The sub-problems are linked by constraints which reference each other. 

\subsubsection{Model I: Instance instantiation and placement}
This model contains constrained variables representing all possible combinations of component and template instances on hosts. Each variable has a discrete domain of \emph{0,1} with \emph{1} indicating presence of an instance. For example, the model for the \emph{Maths} service includes a constrained variable corresponding to host \emph{h3}, component type \emph{MathsService} and component count 4. If, in a solution, this variable has the value \emph{1}, this indicates that the deployed application should locate 4 instances of \emph{MathsService} on host \emph{h3}.

\subsubsection{Model II: Interconnection topology}
This model contains constrained variables representing all possible connections between component instances. For example, the model for the \emph{Maths} service includes a constrained variable corresponding to the provided interface \emph{IMultiplicationService} of the second instance of \emph{MultiplicationService} on host \emph{h2} and the required interface \emph{multiplication} of the third instance of \emph{MathsService} on host \emph{h5}. If, in a solution, this variable has the value \emph{1}, this indicates that a connection should be established between the corresponding component instances.

To improve solving efficiency, some basic \emph{pruning} is performed to avoid creating variables corresponding to impossible connections, for example where interfaces are incompatible. A number of default constraints are placed on the variables in this model, some of which also reference variables in Model I. These  capture the following:
\begin{itemize}
\item{a connection between two component instances may only exist in a configuration if both component instances also exist}
\item{each of a component's required interfaces must be connected to exactly one provided interface}
\end{itemize}

\subsubsection{API}
The API is intended to bridge between the problem-independent general CSP and the problem-specific CSP generated from the DSD. It provides a set of abstractions modelling entities in the DSD's universe of discourse and a set of functions, operating over those abstractions, modeling Deladas functionality. Thus the API provides the abstractions \emph{hosts}, \emph{components} and \emph{templates}. It also supports the abstractions \emph{potential-instance}, modeling a component instance that may be placed on a host, and \emph{potential-connection}, modeling a connection which may or may not exist between two components. Each abstraction supports appropriate operations---for example, \emph{potential-instance} permits the host, component and template type of the potential component it represents to be discovered.

The API functions permit sets of entities to be constructed, manipulated and constrained. The use of the API can be illustrated through the use of a simple example, utilizing four functions from the API: \emph{components()}, \emph{card()}, \emph{lessThanEquals()} and \emph{addConstraint()}. The \emph{components()} function accepts one parameter of type \emph{Host} and returns the set of all \emph{potential-instances} which may exist on that host in some configuration. The \emph{card()} function accepts one parameter which is a set of \emph{potential-instances} and returns an \emph{expression} which represents the cardinality of the provided set. The \emph{lessThanEquals()} function returns a \emph{constraint} which constrains the value of an \emph{expression} to be less than or equal to a given integer. The \emph{addConstraint()} function registers the provided constraint with the solver. 

Using these functions a simple \emph{constraint set} written in Deladas may be compiled into a problem-specific CSP extending the general CSP provided by the framework. The following Deladas example, specifying that the host \emph{h1} should have at most two components placed on it for execution:

\begin{verbatim}
// Deladas specification
constraintSet structuralCons (
  card(components(h1)) <=  2
)
\end{verbatim}
may be transformed into the following Java code:
\begin{verbatim}

// Java CSP specification
public class SpecializedCSP extends GeneralCSP { 
  protected void structuralCons() {
    addConstraint(lessThanEquals(card(components(
      getHostByName("h1"))),2));
  }
}
\end{verbatim}
The generated Java class \emph{SpecializedCSP} extends the class \emph{GeneralCSP}, which implements the functions in the Deladas API, giving the generated code access to the API functions through Java inheritance. The host \emph{h1} referred to in the Deladas specification is manifested in the generated code by a look-up which extracts a set of \emph{potential-instance} variables that may exist on host \emph{h1} from the constraint model. The constraints operate over this set of \emph{potential-instances}. The \emph{structuralCons()} method adds a single constraint to the solver expressing the constraint expressed in Deladas. The API functions mirror those found in the Deladas specification and manipulate the data structures contained in the model to achieve the appropriate semantics. This syntactic trick enables bridging from the Deladas domain to the low-level domain of the constraint solver. 

When \emph{SpecializedCSP} is invoked, and some solution is found, the components assigned to the host \emph{h1} may be determined by calling the \emph{components()} function described above. Configuration details may be extracted in a similar manner from the other types of variables. Thus the assignments made in the solver's domain may be translated back into the high-level domain of the Deladas specification and a configuration description created.

\subsubsection{CDD generation}

The solving framework generates a new CSP for each DSD. The generated classes are dynamically loaded into the solving framework and instantiated. The solver is then invoked to yield a set of solutions for the generated CSP. As described above, solutions to the CSP must be mapped back into the problem domain to yield architecture configurations. This is achieved by iterating over each constrained variable in Model I \& II that represents a potential instance or potential connection. The assignment of the value of 1 to such a variable indicates that an instance or connection should be manifested in the deployment. Architecture configurations are represented as XML \emph{Configuration Description Documents (CDD)}.

\subsubsection{Picking a configuration}

Once a set of candidate CDDs has been generated, one must be chosen and \emph{enacted}. If the user has specified an optimization function, the generated configurations will already be in order of user preference. One possible action for the picker is to simply choose to enact the first and most preferred configuration. However, it is possible that a better decision can be made---for example, if the application is already deployed and the solver has been invoked in response to a DSD violation arising from a host failure. In this case, the first candidate configuration might require that the existing application deployment be considerably evolved. A better choice would be for the picker to select a candidate configuration which is closer to the current deployment, reducing the re-deployment required. The picker is configurable with respect to such policies. 

\subsubsection{Enacting a chosen configuration}

\emph{Enactment} is the process of taking a chosen CDD and creating a running deployment. Enactment of a CDD is a complex task \cite{cingal:framework}, requiring the following:
\begin{itemize}
	\item{Each component is packaged for transmission.}
	\item{Each component package is signed and sent to the appropriate hosts for execution.}
	\item{Upon arrival at a host, each component is verified to ensure that the signer has the appropriate rights to execute a component on that host.}
	\item{Once verified, each component is instantiated in its own address space\footnote{For the purposes of this paper we assume that each component executes in its own address space---however, it is possible to deploy multiple components in a single address space with our implementation.}}.
	\item{Each component is provided with its required interfaces.}
\end{itemize}

Once a CDD has been chosen, the components it describes must be deployed onto the appropriate hosts. This requires that the hosts have the appropriate infrastructure already deployed on them, and that the components are appropriately packaged to support life-cycle management.

To provide a dynamic deployment capability, we have developed a \emph{thin server} architecture \cite{cingal:architecture}, which is used as a hosting platform for dynamically deploying components. Each thin server is essentially a light-weight daemon supporting a single operation \emph{fire()}, which permits a bundle to be instantiated on it, provided that appropriate credentials are presented.

In order to facilitate deployment, each component is packaged into a bundle which may be sent to a thin server. To support life-cycle management, the bundle includes a \emph{Component Manager}, which encapsulates the functionality required to perform deployment, undeployment, monitoring, initialization, destruction, querying of properties, and injection and substitution of service references required by the component. This management functionality, exposed through a set of standard interfaces, is used to monitor and manage each component and thus collectively, the entire deployment.

\subsubsection{Life-cycle management}

In order to execute correctly, components must be provided with references to the services on which they depend. The Deladas Runtime injects such references into components using setter injection \cite{fowler:inversion}. To facilitate this, all inter-component references are made via an automatically generated \emph{smart-proxy} located in the same address space as the caller. If a component attempts to use a service before a reference has been bound, or during configuration evolution, the call blocks in the smart-proxy until a reference is injected. This mechanism decouples the management of the distributed system from the management of an individual component. The benefits are that components are not aware of which other components they are bound to, and that bindings between executing components can be dynamically evolved.

The \emph{Component Manager} carries out a number of actions when instantiating its component:
\begin{itemize}
	\item{it creates appropriate instances of the component's classes, as prescribed in the DSD;}
	\item{it instantiates and records a smart-proxy for each of the required services;}
	\item{it deploys each of the component's provided services to make them available to external clients;}
	\item{if the DSD specifies one or more initialization methods, it calls them to run the component}.
\end{itemize}
 
When a \emph{Component Manager} is requested to undeploy its component, it first disables each of the component's provided services, allowing any ongoing requests to complete. The component's required services are disabled via their smart-proxies, preventing further outgoing calls from being made. Next, it calls the component's \emph{destroy life-cycle methods} declared in the DSD, to allow any necessary cleanup actions to be performed. The final action is to indicate to the thin server that it may terminate any processes required by the component.

\section{Solver Performance}
\label{sec:analysis}

Table~\ref{table:performance} shows the performance of the solver on a variety of  DSDs expressed in Deladas. Performance data was gathered on a single 3GHz Pentium 4 workstation with 1GB RAM running Windows XP (SP2). Each DSD was compiled and solved three times to produce an average of the time required to find all solutions. The times indicated are the CPU time for the Deladas compiler, as reported by the class \emph{java.lang.management.ThreadMXBean}.

\begin{table}[ht]
\begin{center}
\begin{tabular}{| c | c | c | c | c | c | c | c | c | c |}
\hline
DSD &Comp.&Hosts&CSP&CSP&CSP&First&1000&All\\
	&Types		&	    &Variables& Cons. &Solutions&Solution(s)&Solutions(s)&Solutions(s)\\
\hline
1	&	1 		&	1 &  1  		&  1 	& 2 &1.8& - & 1.8 \\ 
2     &     1     		&     2 &  2  		&  2		& 4&1.6 &    -   & 1.6 \\ 
3     &     1            &     4 &  4            &  4        & 16 &2.0&   -  & 2.0 \\ 
4     &     1            &     8 &  8            &  8        & 256 &1.7&  - &  1.7 \\ 
5     &     1            &     16& 16          &  16       & 65,536 &1.7 &1.8& 2.0 \\
6     &     2            &     4  & 80	     &    152      & 123,763,041   &2.1& 2.1 & 1,200 \\ 
7     &     2            &     4  & 80       &    156      & 104     & 2.3 & - & 2.3 \\ 
8     &     2            &     16  & 288   &    592      & $>$180,000,000 & 2.4 & 2.5 & $>$1,800 \\  
9     &     2            &     128 & 16,650 & 33,408 & - & 3.7 &6.9& - \\ 
10   &     2            &     512 & 263,168 & 527,872 & - & 41.0 &85.2 & -\\ 
11   &     3            &     10   & 230   &  481   &  5,634,300 & 2.9 &2.9& 76.7\\
\hline
\end{tabular}
\end{center}
\caption{Solver Performance}
\label{table:performance}
\end{table}
Column 1 contains the experiment number. Columns 2 and 3 contain the number of component types and hosts in the DSD. Columns 4 and 5 show the number of low-level variables and constraints contained in the generated CSP. Column 6 shows the number of different solutions to the CSP found by the solver. Columns 7-9 show the times to find the first solution, the first 1000 solutions, and all solutions.

The table shows the results for four groups of experiments, each demonstrating different facets of the solver's performance. In experiments 1-5, solutions for the deployment of a single instance of a single component type are found, with the number of hosts ranging from 1 to 16. Since every variable in the CSP has a domain of \emph{0,1}, the solution space is exponential in the number of variables, as demonstrated in the numbers of candidate solutions found. 

The effect of constraints on the solution space is demonstrated in experiments 6 and 7. These describe deployments containing client and server components, in which every client requires a single interface provided by a server. In experiment 6 the DSD permits up to two instances of every component type to be deployed on each host. Experiment 7 adds a further constraint that at most one component may be placed on any host. The additional constraint results in a very large reduction in the solution space, and consequently the time taken to find all solutions.

Experiments 8, 9 and 10 show the efficacy of the solver for larger numbers of hosts. From this it can be observed that it is impractical to wait for \emph{all} candidate solutions to be found. However, in every case a first solution is found reasonably quickly. It may not be necessary to wait for all solutions to a DSD in order to receive an \emph{acceptable} configuration. The use of \emph{optimization functions} permits an administrator or a process to specify desirable aspects of a deployment. We anticipate that this will reduce the number of solutions a solver finds and increase the \emph{quality} of those solutions. We intend to investigate mechanisms to trade off the number of solutions generated with the time taken to produce those solutions. One promising policy is to use optimization functions to specify the qualities of a \emph{good} solution, and to specify an upper bound on search time---and select the best candidate configuration found within this bound. This work is ongoing.

Experiment 11 shows the time to solve the \emph{Maths} service example. Despite having a realistic number of component types, hosts and constraints, 5 million solutions are found in a little over a minute.

\section{Status and Future Work}
\label{sec:futurework}
The implementation of our approach is in progress. As of June 2008, we have implemented the language parser, solvers, code generators, and deployment mechanisms. Consequently, we can can deploy distributed (Java) applications. The deployed code includes the component managers, probes and smart-proxies described above.

We are concerned about the number of solutions that are generated by the constraint solver. This may be addressed by adding optimization functions to the DSD, permitting users to specify desirable characteristics of the deployment. Optimization functions permit the constraint solver's search space to be reduced and allow the ranking of solutions.

We are currently building a distributed monitoring infrastructure to collect probe data and events occurring in the deployment. Our current framework has a set of standard probes which monitor aspects of a deployment such as component life-cycle (failure/shutdown), host resource levels and host failure. Assertion probes embodying constraints in the DSD ensure that assertions and constraints hold for the component
or host they are monitoring. Such events will be reported to a \emph{Realm Manager} responsible for monitoring and managing components and hosts. We observe that the runtime detection of constraint violations is a different activity to constraint solving. In the runtime currently under construction, the detection of violations will be the responsibility of the \emph{Realm Manager}, which will dynamically invoke the constraint solver to determine new configurations as shown in Figure~\ref{fig:generalapproach}.

The utility of a configuration may be characterised both in terms of its fit for purpose at some instant and its ability to continue to perform well in the future. The latter may be expressed in terms of \emph{robustness} and \emph{stability}. A robust solution is one that requires minimal alterations in deployment in the face of changes in workload, server failure, network congestion, etc. The nature of these changes may be more or less understood depending on the nature of the deployment environment. Understanding the robustness of a solution gives a measure of the flexibility of the solution. Stability characterises the trade-off between the benefit of deploying a solution and the overall cost of deployment. In some cases, a (re)deployment may deliver high value, while in others the benefit may be outweighed by the cost.

In order for a solver to take account of stability, it would need to model both the cost of enacting a candidate configuration, and the benefit gained by doing so. To take account of robustness, the solver would need to be able to predict the likelihood of future events, and be able to model the impact of the reconfigurations required by those events.

\section{Conclusions}
\label{sec:conclusions}

We believe that automatic management of distributed application deployment will become essential as the scale and complexity of applications grow. This paper has outlined a framework to support the initial deployment and subsequent evolution of distributed applications in the face of perturbations such as host and link failure, temporary bandwidth problems, etc. The knowledge required for automatic management is specified in the form of a set of available hardware and software resources and sets of constraints over their deployment.

We have demonstrated that it is possible to apply \emph{constraint satisfaction} techniques to the problem of finding deployments that are compliant with a declarative specification. The approach we have taken involves the creation of a generic constraint satisfaction problem which describes the general problem of component placement and component interconnection. This general problem is specialised by generating a domain-specific constraint problem from the specification. We have shown that constraint solvers can produce solutions quickly even in cases where the number of potential solutions are extremely high. We hypothesise that when optimisation functions are used the solution space will be considerably reduced.

We have also described how components can be generated that contain the infrastructure needed to control the entire component life-cycle including instantiation, destruction, monitoring and evolution. This is achieved by incorporating a \emph{Component Manager} along with non-invasive smart-proxies into the deployed code. The smart-proxies permit inter-component bindings to be changed within an executing components and hence the distributed application to be evolved.

We have sketched how constraint satisfaction may be used at run-time to control the evolution of applications. We are actively working on the next stage of implementation which will support constraint-led evolution of distributed applications and hope to be able to report on this in the near future.

\section{Acknowledgements}

This work was supported by EPSRC Grants GR/M78403 Supporting Internet Computation in Arbitrary Geographical Locations, GR/R51872 Reflective Application Framework for Distributed Architectures and EP/C014782/1 Design, Implementation and Adaptation of Sensor Networks through Multi-dimensional Co-design.

\bibliography{bibliography}
\end{document}